\documentclass[10pt,journal, compsoc]{IEEEtran}

\usepackage[utf8]{inputenc}
\usepackage{url}
\usepackage{niravstyle}
\usepackage{graphicx}
\usepackage{import}
\usepackage{multirow}
\usepackage{multicol}
\usepackage{rotating}

\usepackage{booktabs}

\usepackage{siunitx}
\usepackage{amssymb}
\usepackage{amsmath}

\newcommand{\aod}[1]{\textcolor{blue!50!black}{AOD:~~#1}}
\newcommand{\revision}[1]{\textcolor{magenta!50!black}{#1}}

\usepackage[numbers]{natbib}

\newtheorem{definition}{Definition}

\newcommand{\rx}{\rotatebox{60}}
\newcommand{\ftx}{\textit}
\renewcommand{\fbf}{\bfseries}

\renewcommand{\nsa}[1]{}
\renewcommand{\add}[1]{}
\renewcommand{\del}[1]{}
{}
\renewcommand{\aod}[1]{}
\renewcommand{\revision}[1]{}

\usepackage{siunitx}
\usepackage{enumitem}
\usepackage[inline]{enumitem}

\usepackage[ruled,vlined,linesnumbered]{algorithm2e}

\author{Alex Davies, Nirav Ajmeri}

\begin{document}

\title{Realistic Synthetic Social Networks with Graph Neural Networks}

\IEEEtitleabstractindextext{
\begin{abstract}


Social network analysis faces profound difficulties in sharing data between researchers due to privacy and security concerns.
A potential remedy to this issue are synthetic networks, that closely resemble their real counterparts, but can be freely distributed.
generating synthetic networks requires the creation of network topologies that, in application, function as realistically as possible.
Widely applied models are currently rule-based and can struggle to reproduce structural dynamics.
Lead by recent developments in Graph Neural Network (GNN) models for network generation we evaluate the potential of GNNs for synthetic social networks.
Our GNN use is specifically within a reasonable use-case and includes empirical evaluation using Maximum Mean Discrepancy (MMD).
We include social network specific measurements which allow evaluation of how realistically synthetic networks behave in typical social network analysis applications.

We find that the Gated Recurrent Attention Network (GRAN) extends well to social networks, and in comparison to a benchmark popular rule-based generation Recursive-MATrix (R-MAT) method, is better able to replicate realistic structural dynamics.
We find that GRAN is more computationally costly than R-MAT, but is not excessively costly to employ, so would be effective for researchers seeking to create datasets of synthetic social networks.
\end{abstract}

\begin{IEEEkeywords}
    Graph Neural Network, Social networking (online), Graphs and Networks.
\end{IEEEkeywords}
}


\maketitle

\revision{Hello in magenta}

\section{Introduction}
\label{sec:introduction}




Social network data is widely used across a variety of domains. This includes in political research to model how different groups vote \cite{Abdine2022PoliticalElection}, in researching conflict between communities \cite{Kumar2018CommunityWeb}, in modelling how we spread rumours online \cite{Indu2019ANetworks}, tracking COVID in the earliest stages of the pandemic \cite{Yin2020COVID-19Sina-microblog}, how reviews influence what we buy \cite{Xiong2020ExploitingRecommendation}, and to predict voter behaviours \cite{Khatua2020PredictingContexts}. An overview of social network analysis applications can be found in \citet{Can2019AApplications}. 

Despite the level of interest in studying social networks researchers often cannot easily share their data. The obstacles to sharing data are primarily privacy and security, which is particularly is of concern where data focuses on individuals. In extreme cases, researchers may not be able to access the original data at all. 

There is a wealth of work on anonymising social network data, where a network is altered to preserve the anonymity or the security of the information it contains, so that it can be shared and used more freely.
\citet{Yuan2013ProtectingAnonymization} propose anonymisation through addition of ``noise nodes'',  \citet{Jian2021PublishingPrivacy} propose anonymisation under node and edge differential privacy, and \citet{ZhouBin2008AData} provide a brief survey of anonymisation approaches.
While there is extensive work on anonymising social network data, there is a near-equal level of work on \textit{de}-anonymising previously anonymised social networks.
\citet{Narayanan2009De-anonymizingNetworks},  \citet{Fu2020De-AnonymizingStructure}, \citet{Caragiannis2019DeanonymizingInformation} and  
all demonstrate that with a combination of network data and limited user data individual identities can be re-constructed. 
In particular these works make use of overlapping networks and information between datasets, effectively cross-referencing to narrow the scope of possible user identities.
This means that even bare topologies---without any user or connection information---can represent, in combination with other data, a privacy or security risk.




To this end, methods of creating synthetic networks have been developed, aimed at allowing the sharing of data, while alleviating privacy or security concerns. Currently employed methods are generally rule-based \cite{Chakrabarti2004R-MAT:Mining}. The ability to generate realistic networks  would allow far easier sharing of data between researchers without the issues that follow using real data. Commonly employed rule-based methods can struggle to generate networks with structure that mirrors their real counterparts beyond node-level distributions. 


Meanwhile, following the advent of Message Passing Neural Networks (MPNNs) \cite{Gilmer2017NeuralChemistry}, there has been the recent development of powerful Graph Neural Network (GNN) models. Having been used for social network analysis tasks, when turned to generation in other fields, they have shown an ability to generate complex and meaningful structure \cite{DeCao2018MolGAN:Graphs}. This means that application to social network analysis, where topology plays a significant role, is promising.

\subsection{Motivating Use-Case}\label{sec:usecase}


We present as motivation the scenario of a researcher, who we name Mark, who works at a social media company.
Mark has a dataset of relatively small networks.
These networks, due to their nature, cannot be distributed verbatim without risking the privacy or the security of the individuals the data concerns. 
These networks could be from social media, internal communications from Mark's company, inter-community linking on a platform like Reddit, or any other medium where the anonymity of those involved must be preserved.
The options presented to Mark in order to share their data are:

\begin{description}[nosep,leftmargin=0em]
    \item[Alter the topology.] Several methods exist (e.g.\add{,} edge swapping) to alter networks in order to anonymise them. These require changing, to some extent, the nature of the networks.
    \item[Use rule-based synthetic topologies.] Use a method that follows explicit steps and parameters to create synthetic versions of the networks.
    \item[Use GNN generated synthetic topologies.] A new option, and one which hasn't been tested, is to use a GNN model to generate synthetic versions of the networks.
\end{description}

Mark has computational and time constraints in how they can produce their distribute-able dataset. In order for GNN generated topologies to be the superior option from these three, Mark has to consider:

\begin{description}[nosep,leftmargin=0em]
    \item[The similarity of networks] The generated network topologies should correspond as closely as possible to the distributions underlying the original dataset.
    \item[The computational expense] Rule-based generation and topology alteration are comparatively cheap operations, so a GNN model should not be excessively costly to employ
    \item[The time to use] The run-time of fitting a model and generating topologies should not be prohibitive
\end{description}

\subsection{Research Questions}


This work aims to address these factors that Mark must consider.
We envisage that with GNN methods, as in other fields of deep-learning, there is a compromise between superior results and computational expense, in comparison to existing methods. Explicitly, our research questions are:


\begin{enumerate}[label=\textbf{RQ\arabic*}]
    \item \textbf{How could we generate synthetic social network topologies comparable to real networks?}
    
    
    To address this question we benchmark a GNN model against a popular rule-based model in the task of generating synthetic social networks.
    Here we use GRAN \cite{Liao2019EfficientNetworks} and R-MAT \cite{Chakrabarti2004R-MAT:Mining} for GNN and rule-based methods respectively.
    Comparison is performed using both empirical analysis via Maximum Mean Discrepancy (MMD) and qualitatively via visual inspection. \label{rq1}
    
    
    \item \textbf{When are GNN models more useful for generating synthetic datasets than the current standard?}
    
    To address this question we compare, using social network specific metrics and visual inspection, networks created by GRAN to those from the rule-based method R-MAT \cite{Chakrabarti2004R-MAT:Mining}. \label{rq2}
    
    \begin{enumerate}[label*=\textbf{.\arabic*}]
        \item \textbf{What is the trade-off between computational cost and ability to create realistic social networks?}
        
        To address this question we compare the computational and time investment required when addressing the previous questions.\label{rq3} 
  \end{enumerate}
\end{enumerate}




We place particular emphasis on domain-specific applications via shortest path lengths, SIR simulation and Louvain community detection \cite{DeMeo2011GeneralizedNetworks}.




\subsection{Organisation}



Section \ref{sec:related} details the relevant work on synthetic social network generation, which includes work from other fields on graph generation.
Section \ref{sec:preliminaries} details information essential to understand and critique this work.
This section includes technical background on models employed in this work.
Section \ref{sec:methodology} details our methodology, \aod{... update here based on what ends up being included} the scenarios in which our methodology would be used, as well as how our method is used in those scenarios.

Section \ref{sec:experiments} consists of evaluating how a GNN model fairs at generating synthetic social networks, including quantitative analysis via MMD, and qualitative analysis via visual inspection.
Also included in Section \ref{sec:experiments} is a discussion of computational and time costs.

Section \ref{sec:discussion} details the key findings of the experiments in Section \ref{sec:experiments}, structured according to our research questions, along with a discussion of assumptions or simplifications we've made, as well as potential risks to the validity of our findings.
Finally Section \ref{sec:conclusion} gives a high level view of this work, with condensed findings and a high-level summary of our method.
Section \ref{sec:conclusion} also gives a summary of the results of our experiments with GRAN and R-MAT.

\section{Background and Related Work}
\label{sec:related}



Our work is most closely related to graph generation. 
Graph generation presents complexities not present in other domains of generative models \cite{Zhang2022DeepSurvey}, as models and algorithms must be resilient to variable node orderings.
Additionally, there is no ``standard'' sizing of networks as there is with, for example, image resolutions. 
Graph generation is also often used in domains where the size of the network should be scalable to an arbitrary number of nodes, for example, in molecule generation \cite{DeCao2018MolGAN:Graphs}.
\citet{Zhao2020AGeneration} give an overview of network generation.
Alternative reviews, including example applications, can be found in \citet{Zhu2022AApplications}~and~\citet{Zhang2022DeepSurvey}.
Though newer, these reviews do not cover some successful auto-regressive models.



\subsection{Rule-Based Methods}

Human networks often have the ``small world'' characteristic, where the average distance between nodes is small and consistent across networks sizes, as demonstrated in the six degrees of freedom experiment \cite{TRAVERS1977AnProblem}.
\citet{Watts1998CollectiveNetworks} propose a random generator that exhibits these desired properties: high clustering coefficients and small diameters.
This is the most simple form of the rule-based systems.
Each node has a set number of neighbours and connections between nodes are randomly sampled.

This presents issues in that each node retains the same number of neighbours, leading to a lack of divergence from starting conditions, resulting in networks that can be over-simple or non-realistic.
Real-world networks present a power-law distribution in node degrees.
A more recent generative method, R-MAT \cite{Chakrabarti2004R-MAT:Mining}, generates via a recursive division of the original network's adjacency matrix, and results in networks with this power-law distribution.

R-MAT remains commonly used in recent works, often adapted with changes to domain-specific applications \cite{Nettleton2021MEDICI:Generator, Gursoy2019AGeneration, Nettleton2016AGraphs}.
This includes the distribution of node-level attributes across topologies generated by R-MAT.
Notably, all of these methods are recursive or rule-based, and may miss more complex dynamics of social networks, in particular in generating meaningful large-scale structure. 

\subsection{Deep-Learning Methods}

While synthetic social networks aren't new, with algorithms and models such as R-MAT \cite{Chakrabarti2004R-MAT:Mining} already fairly popular, deep-learning models have not widely been applied to real social network data.
The recent development of Message Passing Neural Networks, arguably proposed by multiple papers like \citet{Gilmer2017NeuralChemistry}, has allowed rapid development of deep-learning models for network data structures.
This includes the development of algorithms for network generation, with much recent work in chemistry. 

Deep-learning approaches to network generation vary considerably.
Some approaches model network construction as biased random walks, such as NetGAN \cite{Bojchevski2018NetGAN:Walks}.
Others are ``one-shot'' methods, where advancements in fields like machine-vision are transferred to the network generation problem, for example GraphGAN \cite{Wang2017GraphGAN:Nets}.
Often models build on architectures from other forms of data, like images, with papers such as  \citet{DeCao2018MolGAN:Graphs} and \citet{Simonovsky2018GraphVAE:Autoencoders} adapting GANs and VAEs.
These models are limited to small networks due to issues presented with node orderings and the rapidly scaling memory requirements of networks with increasing size. 


Only even more recently have models such as GRAN \cite{Liao2019EfficientNetworks} and TG-GAN \cite{Zhang2021TG-GAN:Constraints} been able to generate networks of the sizes found in social network analysis.
The lack of application of models such as GRAN and TG-GAN on social network data specifically is what this work aims to address.
Other works in network generation have not evaluated deep-learning models in comparison to rule-based methods, whereas in generating synthetic social networks, rule-based methods are the current benchmark.
Additionally, other works have not considered metrics specific to social networks, which this work also aims to address.




\section{Preliminaries}
\label{sec:preliminaries}


Here we describe the preliminaries necessary to understand and critique our method and results. 






\subsection{Network Structures \& Terminology}

    

We begin by defining the terminology and notation that we use throughout the rest of this work.

\begin{definition}[Network]
    \normalfont
    Networks, or in more abstract terms graphs, are relational structures of data. They are generally collections of the form ``$A \rightarrow B$''. We denote a network as $G$, and a set of networks as $\mathbb{G} = \{G_1, G_2,...\}$.
\end{definition}

\begin{definition}[Un-directed network]
    \normalfont
     In an un-directed network any given connection is mutual, $A \leftrightarrow B$.
\end{definition}

\begin{definition}[Nodes, Node-set]
    \normalfont
    In social networks each node might be a user, a page, a website, and so on. 
    The nodes in a network are described as a node-set $V: \{v_1, v_2, v_3, ...\}$.
\end{definition}

\begin{definition}[Edges, Edgelist]
    \normalfont
    Edges between nodes in a social network might be friendships, collaborations, shared interests, or some other relation.
     The edges between nodes are contained in an edgelist, where an individual edge is $(v_i, v_j)$ for a connection $v_i \leftrightarrow v_j$, and so an edgelist has the form $E: \{ (v_i, v_j), (v_k, v_l), (v_m, v_o),... \}$. 
\end{definition}

\begin{definition}[Neighbours, Degree]
    \normalfont
    The neighbours of a node in an un-directed network are the other nodes which a node is connected to, ie $N(v_i) = \{v_1, v_2, v_3,...\}$ if there are edges $\{(v_i, v_1), (v_i, v_2), (v_i, v_3),...\} \in E$.
    In an un-directed network the degree of a node is the size of its neighbourhood, $d(v_i) = |N(v_i)|$
\end{definition}

\begin{definition}[Shortest Path]
    \normalfont
    The shortest possible route (in terms of number of edges) between a given node pair, $SP(v_i, v_j) = \{ (v_i, v_1), (v_1, v_2), ...,(v_n, v_j),\}$.
\end{definition}

\begin{definition}[Connected Component]
    \normalfont
    A connected component is a network or sub-network in which there is a possible path between any given pair of nodes.
\end{definition}

\begin{definition}[Motif]
    \normalfont
    A motif is a generally small sub-network which may occur more than once in a larger network. We denote a motif as $C$.
\end{definition}





\subsection{GRAN}\label{sec:gran}

A recent and successful network generation model is the Graph Recurrent Attention Network (GRAN) \cite{Liao2019EfficientNetworks}. For temporal networks, a successful model is TG-GAN \cite{Zhang2021TG-GAN:Constraints}, which aims to encapsulate the dynamics of networks with changing topology. We focus on the slightly simpler GRAN model, as evaluating the quality of synthetic topologies is less opaque without accounting for temporal changes.

GRAN is a ``graph-motif'' auto-regressive model, meaning that it adds nodes and edges in blocks, with each motif at step $t$ denoted $C_t$. The primary issue that GRAN aims to solve is where to add new motifs $C_t$ to the previously constructed sub-network $G_{t-1}$.

For a generative process with motif size $B = |C_i|$, at step $t$, there are $B(t-1)$ nodes in the current network.
At step $t$ we add $B$ new nodes, and fully connect them to the existing network.
Nodes are initialised with a node-level hidden state of $h_{n,i} = 0$.
From here a Message-Passing Neural Network (MPNN, \cite{Gilmer2017NeuralChemistry}) updates the hidden states of nodes. 
After several rounds of message passing, node states are passed to an MLP to calculate a Bernoulli distribution for edge existence. 
MLP$_{\theta}$ parameterises the Bernoulli distribution for the edge existence between $i$ and $j$.
The resulting network is then used as input for the next block $C_{t+1}$, and so the model generates auto-regressively.







\subsection{Evaluating Generated Networks}




In network generation, as in other areas of generation, measurement of quality can represent a non-negligible challenge. In image or text generation, which have seen significant research, evaluation can be performed visually or semantically, especially when generated content quality is particularly poor. In network generation visual inspection is via layout algorithms, which may introduce their own biases, and such visualisations can remain relatively opaque with networks of many nodes. 

Unlike other domains of generation, such as images, MAE/MSE or some other direct metric does not apply.
Graph metrics generally describe either the network as a whole, or a node within a network, respectively termed network-level and node-level metrics.



Commonly, papers proposing network generation methods make some measurement of the network, then compare the result to the original dataset.
This can be as simple as comparing mean values \cite{Ali2014SyntheticData}, or more expressively make use of Maximum Mean Discrepancy (MMD) to allow variation across the dataset, which is a kernel-based method for comparing distributional similarity.
Notably the metrics most often applied are not domain specific.
We propose that demonstrating the quality of synthetic social networks requires consideration of their use in social network analysis.
To this end we employ common algorithms: SIR simulation to model information movement across the network, Louvain community detection, and shortest path lengths.
These are aimed at assessing the relative quality of the structure in synthetic topologies.




\section{Methodology}
\label{sec:methodology}

\aod{Come back to this - going to be a lot of changes}

\aod{Should include explicit data prep steps? Making files etc? Not normally included in DL papers}

The use-case we model is that of a researcher with a dataset of (relatively) small networks, who wishes to publish a dataset of realistic synthetic versions of the same. This allows assessment of generated topology structures while remaining within the constraints of the model we're using, and without excessive computational run-time.





\subsection{Sampling \& ESWR}
\label{sec:eswr}

In order to construct a network dataset for experimentation, we propose Exploration Sampling With Replacement (ESWR), which takes as input a single large network, and outputs a dataset of sampled sub-networks. In other machine learning applications SWR is not advised, as models tend to overfit if they are presented the same data more than once. In this application there are many degeneracies in how networks are presented to a deep-learning model, primarily node ordering. This means that while overlapping sections of the overall network may be presented to the model as separate inputs, the model is unlikely to overfit to these sections. We argue that ESWR presents several advantages:

\begin{description}
    \item[Representative samples] Exploration sampling algorithms, for example diffusion sampling, are specifically designed to sample a network such that a given sample exhibits the same characteristics as the original network. This is not true of more simple methods like random node sampling. 
    \item[Size of dataset] Sampling \textit{without} replacement would, after the first sample, be sampling a network with sections ``missing'', which would mean that samples have decreasing similarity to the original network. ESWR means that, while avoiding this issue, more nodes can be present in the sampled dataset than in the original network.
\end{description}

The application of ESWR to a network is described by Algorithm~\ref{alg:eswr}, here to generate an overall edgelist and a list of network identifiers. Here $\mathcal{N(\mu, \sigma)}$ is a normal distribution of mean $\mu$ and deviation $\sigma$, as a simple of example of a sample-size distribution to use. As previously, $G$ is the input network, $E$ is the list of edges, and $V$ is a set of nodes. Some elements are omitted for the sake of brevity, for example re-indexing nodes for each sampled network.

\begin{algorithm}
$G \gets$ LargeNetwork  \tcp*{Original, $|G| > \mu$} 
$G_{ind} \gets []$\;
$E \gets []$\;
\For{$i \gets 0$ to $N_{networks}$}{
    $|V_{sample}| \gets $SampleSize$()$\; 
    $G_{sample} \gets$ ExplorationSampler$(G, |V_{sample}|)$\;
    $G_{ind} \gets G_{ind} + [i \times |V_{sample}|]$\;
    $E \gets E + E_{sample}$}
\caption{ESWR to construct a dataset of sub-networks from a single large network}
\label{alg:eswr}
\end{algorithm}

\subsection{Model Use}



GNN models for generation typically require a dataset which a model learns to replicate. In order for fair evaluation, as is standard in most deep-learning methods, we split the dataset into training and testing. This isn't necessarily the standard in deep-learning models for generation, as performance is often evaluated visually or semantically, and quantitative metrics are not necessarily a good indicator of quality. In network generation, and in this work, quantitative metrics are more widely used. Given the relevance of quantitative metrics the quality of a generated synthetic dataset is evaluated against networks on which the GNN model has not been trained.

In keeping with our realistic use-case we refer back to Section~\ref{sec:usecase}. We assume that Mark, our model user, has access to moderate computing resources but not to a full HPC facility. Additionally, Mark has limited time to create their synthetic dataset. In these circumstances, which we envision are encountered by most of those employing synthetic datasets, the actual utility of a model is determined not only by the quality of the networks it produces. 

Taking these factors into account, we do not include a full hyper-parameter optimisation process, and also avoid excessive training durations. We emphasise that the aim of this work is to evaluate the \textit{realistic} performance of GNN models in this application, instead of the maximum performance possible with unlimited resources.

\aod{Move to experiments
\subsubsection{Rule-Based Model Use}
As a benchmark for performance, we include comparison to a rule-based method, which are currently employed for synthetic datasets.
This rule-based model should constitute a fair comparison.
This would not include a particularly old or seldom used model.
Similarly the rule-based model should aim to achieve the same goals as the GNN model, or if designed for use in another field, have achieved comparable results in that field.
For evaluation the rule-based model should, as with the GNN model, be applied to the test-set data.
If the rule-based model functions on a per-network basis this should still be done, but with proper consideration of the potential for overfitting.
}

\aod{
\begin{itemize}
    \item Most proposed attacks for de-anonymisation assume a ground truth
    \item Some use neighbourhood attacks (ie assume attacker has 1-hop or 2-hop) neighbourhood \cite{Wang2013OutsourcingCloud}
    \item The risk here is that these motifs can be re-produced by GRAN
    \item Perform an analysis of different neighbourhood sizes in real graph vs occurrences in synthetic networks
\end{itemize}
}
\aod{
\begin{itemize}
    \item My thinking is to randomly sample $N(k)$ networks from the synthetic networks
    \item $N$ is the number of networks
    \item $k$ is the size of each network
    \item Sampling would probably be extended ego networks (eg one-hop, two-hop)
    \item Also test against ESWR sampled mini-networks
    \item \url{https://graph-tool.skewed.de/static/doc/topology.html?highlight=isomorphism#graph_tool.topology.subgraph_isomorphism}
    \item can check for subgraph isomorphism (ie ``is this in the training data'')
\end{itemize}
}
\aod{
\begin{itemize}
    \item Turns out subgraph-isomorphism checks are slow
    \item Very slow for more than a few nodes
    \item Small-size networks are very commonly isomorphic
    \item Large size networks have lower subgraph-isomorphisms are less common
\end{itemize}
}
\section{Experiments}
\label{sec:experiments}

Here we detail the experiments we perform.

\subsection{Data}

We sample several large networks published by \citet{Rozemberczki2019GemSec:Clustering} using ESWR. These are available from the Stanford SNAP repositories \cite{JureLeskovec2014SNAPCollection}. For our exploration sampling algorithm we use Metropolis Hastings Random Walk sampling, implemented in the Little Ball of Fur Python library \cite{Rozemberczki2020LittleSampling}. For each individual sample, a random number of nodes is sampled from the larger network, with the number of nodes sampled from a normal distribution, $\mu = 400, \sigma = 50$. The use of a normal distribution is aimed at giving a model the ability to learn a simple distribution for network size. 

\begin{table}[!htb]
\caption{Details of our large network datasets, sourced from \citet{Rozemberczki2019GemSec:Clustering} and \citet{Rozemberczki2020KarateGraphs}}
\label{tab:datasets}

\centering
    \begin{tabular}{lcccc}
    \toprule
                   & {Facebook} & {GitHub} & {Twitch} & {Deezer}\\\midrule
    N. Networks    & 200                     & 200      & 200      & 9629 \\
    Min. Nodes       & 279                     & 288      & 280      & 11 \\
    Max. Nodes       & 531                     & 525      & 533      & 363 \\
    Min. Density     & 0.00945                 & 0.00621  & 0.00840  & 0.0150 \\
    Max. Density     & 0.0361                  & 0.0138   & 0.0145   & 0.909 \\
    Min. Comm.       & 9                       & 13       & 12       & 1 \\
    Max. Comm.       & 24                      & 21       & 19       & 12 \\\bottomrule
\end{tabular}

\end{table}

200 sub-networks are sampled per large dataset. Table~\ref{tab:datasets} provides details for the resulting datasets after sampling. We also include a larger dataset of ego networks, which is sampled from the music social network site Deezer \footnote{A music streaming service with social network features,~\url{https://www.deezer.com/}}, published by \citet{Rozemberczki2020KarateGraphs}.

\subsubsection{Assessment of Network Characteristics}

Ego networks have the defining property of having a central ``hub'' node to which all other nodes connect. This is also true of the Deezer networks. The nature of these networks might prove difficult to encode and reproduce, as it is semantically complex, and could require modelling beyond a degree distribution in larger networks. The Deezer networks are generally small, so are easy to visualise, making this dataset something of a toy example to examine the potential benefits of using a GNN model. 

For networks sampled from the Facebook network, we expect tight clusters, joined by low degree ``arcs'' of nodes. This represents pages of the same type, or in the same community, with other pages bridging the gaps between these communities. As the number of nodes in a network increases we expect a large range in numbers of communities found by application of Louvain detection. Small Facebook sampled networks will have higher density than those sampled from other social networks, as seen in Table~\ref{tab:datasets}, and a wide range of node degrees.

Graphs sampled from Twitch are similar to those sampled from Facebook but with fewer and larger clusters. As seen from Table~\ref{tab:datasets}, this manifests in a smaller range in the number of communities found by Louvain, and lower minimum and maximum densities. Arcs of low degree chains of nodes are still present, but connect back to a central cluster, instead of joining otherwise separate clusters.

GitHub sampled networks have lower densities than Twitch sampled networks. We can again expect to have a clear central grouping of nodes. Like the Twitch networks, this will have arcs of low degree nodes that connect back to the central grouping. This results, under visualisation, in a ``ball-of-wool'' appearance.

\subsection{GNN Model}


\label{sec:gran_gen}

As an example of current cutting-edge GNN models for network generation we use GRAN \cite{Liao2019EfficientNetworks}. The implementation distributed by \citet{Liao2019EfficientNetworks} has two modes of use: training and testing. Training is used for model training, and requires a dataset (in the form described above) as input, as well as architecture and training related parameters. Testing is used to compare a trained model's generated networks against a real dataset, primarily to produce MMD scores, and also produces visualisations of generated and real networks.

We add functionality, including social-network specific metrics and algorithms for MMD calculation, as well as a ``generate'' mode. This mode does not require input files, and simply employs a trained model to produce visualisations and edgelists of generated networks. Optional configuration, via .yaml files, allows the MMD functionality of the testing mode to be included in generation mode. We also add the ability to generate R-MAT networks, but as these are fit on individual network samples, this is not by default an aspect of generation.


Following the precedent set in the original work's training, we chose GRAN parameters and architecture depth as detailed in Table~\ref{tab:gran_params}.
These parameters are used for each dataset.




\begin{table}[h]
    \centering
    \caption{Parameter selection for GRAN models.}
    \label{tab:gran_params}
    \begin{tabular}{cc|cc}
    \toprule
    Node Order & DFS & Num. Mix. Comp & 20\\
    Block Size & 1   & Sample Stride  & 1 \\
    Hidden Dim.& 512 & Max. Nodes     &550\\
    Embedding Dim. & 512 & GNN Layers & 7\\
    GNN Prop.  & 1 &&\\
    \bottomrule
    \end{tabular}

\end{table}






\subsubsection{Training - Parameters}

We emphasise that our aim is not to reach the limits of performance with GRAN, but to evaluate whether a model such as GRAN can be rapidly deployed by users to create synthetic social networks of a higher quality than those produced by rule-based methods. To this end we aim to constrain resources and computation time to less than a day, which we argue constitutes a rapid deployment, so training is limited to 30000 epochs for each dataset. An exception is the model fit on the Deezer networks, which is trained for 10000 epochs. 


The size in memory of a network varies considerably with the size of the node-set, and further, networks with the same number of nodes may have greatly varying numbers of edges (to a limit of $0 \leq |E| \leq \frac{1}{2}|V|^2 - |V|$). This leads to batches with the same numbers of networks having hugely variable size in memory. To this end we limit batch sizes for the three sampled datasets (Facebook, Twitch, GitHub) to 50. The Deezer networks are smaller, reducing the uncertainty in memory usage, so we are able to set a batch size of 200. As in \citet{Liao2019EfficientNetworks} we take a train/test split of 80:20. For the three datasets constructed through ESWR this means a test set of 40 networks and a train set of 160. For the Deezer dataset the test set is 1925 networks and train set 7703.

\subsubsection{Training - Resources}

Training and model evaluation is conducted on a desktop workstation. Using a workstation, instead of an HPC facility, is done in accordance with our fictional user Mark. We detail the workstation in the Appendix.
While the workstation is powerful compared to consumer-level computers, in the context of deep-learning research where access to HPC facilities can be assumed, it is not excessively so.


\subsection{Benchmarking via R-MAT}
\label{sec:rmat_gen}


R-MAT (Recursive MATrix) is a model proposed by Chakrabarti et. al. in 2004 \cite{Chakrabarti2004R-MAT:Mining}. It functions on the basis of recursively sub-dividing an adjacency matrix of size $|V|$ into equally sized  quadrants. As parameters a user passes the probability of an edge falling into each of these four quadrants, via parameters $a,b,c,d$, trivially $a+b+c+d = 1$. The size of a generated network is fixed to some integer factor of $2$, ie $|V| = 2^n; n = \log_2(|V|)$. 

In this work we use the R-MAT implementation from NetworKit \cite{Staudt2014NetworKit:Analysis}. This takes as input the previously described parameters $a,b,c,d$, and also ``edgeFactor'' $|E| = |V|*\textrm{edgeFactor}$, to determine the number of edges a given network should have. We are able to remove nodes via the parameter ``reduceNodes'' in order to have a non-factor-of-two number of nodes. This allows us to generate networks with a number of nodes matching each testing example.

For comparison between GRAN and R-MAT generated networks, we fit parameters and generate an R-MAT network for each network in a given dataset. More recent rule-based methods have been proposed in other works, but R-MAT remains common in the field, and serves as a good benchmark against which to compare GRAN.

\subsection{Metrics}

We employ common node and network-level metrics, as well as metrics tailored to social network analysis, with quality calculated primarily via Maximum Mean Discrepancy (MMD).
Lower MMD scores indicate more similar distributions with identical distributions scoring $MMD = 0$.


MMD requires the measurement of the distribution of a given quantity. Several are packaged in the implementation of GRAN from \citet{Liao2019EfficientNetworks}, and we add measured quantities specific to social network analysis:


\begin{description}
    \item[MMD\fsub{Nodes}] The number of nodes in each sub-network, $|V|$. We employ a Gaussian kernel to calculate MMD, as we know that the desired node distributions follow a normal distribution. Note that this is in contrast to \citet{Liao2019EfficientNetworks}, who use a Gaussian Total Variance (TV) kernel.
    
    \item[MMD\fsub{Degree}] The degree of each node, denoted $d(v)$ for a node $v$. Calculated using a Gaussian kernel.
    
    \item[MMD\fsub{Clustering}] The fraction of possible triangles through a node that exist i.e. the extent to which a node and its neighbours are a complete network. Clustering $c_v$ for a node $v$ is calculated as 
    \begin{equation}
        c_v = \frac{2T(v)}{d(v)(d(v)-1)}
    \end{equation}
    where $T(v)$ is the number of triangles through $v$, and $d(v)$ is the degree of $v$. Dense clusters in networks have high clustering coefficients, due to their high inter-connectedness, whereas more sparse regions have low clustering coefficients. Calculated using a Gaussian kernel.
    
    \item[MMD\fsub{Spectral}] 
    For an un-directed network $G$, we can construct a diagonal matrix of node degrees $D$. From this we can calculate the Laplacian of the network, $L = D - A$, where $A$ is the network's adjacency matrix. The normalised Laplacian is then:
    
    \begin{equation}
        N = D^{-1/2}LD^{-1/2}
    \end{equation}
    

    Taking the eigenvalues of this normalised Laplacian allows us to treat the network as defined by a spectrum. Spectral analysis takes a view of global network properties, unlike degree or clustering, which are local statistics. We calculate MMD\fsub{Spectral} using these Laplacian eigenvalues and a Gaussian TV kernel.
\end{description}

\begin{description}
    \item[MMD\fsub{Paths}] All node-node shortest path lengths in a network. The shortest path between two nodes is the path that which contains the fewest nodes. Calculated using a Gaussian kernel.
    
    \item[MMD\fsub{Steps}] The number of steps an SIR simulation on a network takes before termination. We implement a simple SIR (Susceptible, Infected, Recovered) model. All nodes begin as ``Susceptible'', then a randomly selected subset of size $N = 2$ are set as ``Infected''. Infected nodes at each iteration have a $\kappa = 0.04$ probability of infecting each of their Susceptible neighbours. Once a node has been Infected for $\gamma = 5$ iterations it becomes Recovered, can no longer infect its neighbours, and in our implementation, also cannot become reinfected. We measure the number of iterations until no further infection is possible, or until 100 iterations have taken place. This is simulation is performed multiple ($n=20$) times to account for the random selection of seed nodes. Calculated using a Gaussian kernel.
    
    \item[MMD\fsub{Saturation}] The proportion of nodes in a network that are in a ``Recovered'' state at the termination point of an SIR simulation. Calculated using a Gaussian kernel.
    
    \item[MMD\fsub{Louvain}] The number of communities found by application of Louvain community detection. Louvain community detection is a modularity-optimising, un-supervised algorithm for finding communities in networks. We use the implementation from the Python library Python-Louvain \cite{Blondel2008FastNetworks}, with default parameters, ie with a resolution of $1$. Again we run the algorithm multiple times for each sub-network to account for random elements of the algorithm, but given the increased stability of Louvain compared to SIR simulations, we perform only $n=5$ runs. Calculated using a Gaussian kernel.
    
\end{description}


\subsection{Results}

Here we detail the results of our experiments.

\subsubsection{Quality of Topologies - Empirical}
\label{sec:empirical}

\begin{table*}[!htb]
    \centering
    \caption{Measurements of the test set networks and networks generated by GRAN for each of our datasets.}
    \label{tab:gen_stats}
    \begin{tabular}{l SS SS SS SS}
    \toprule
         & \multicolumn{2}{c}{Facebook}  & \multicolumn{2}{c}{GitHub}   & \multicolumn{2}{c}{Twitch}  & \multicolumn{2}{c}{Deezer} \\
         \cmidrule(lr){2-3} \cmidrule(lr){4-5} \cmidrule(lr){6-7} \cmidrule(lr){8-9}
         & {Real} & {GRAN} & {Real} & {GRAN} & {Real} & {GRAN} & {Real} & {GRAN} \\\midrule
        Min. Nodes & 283.000000 & 183.000000 & 294.000000 & 288.000000 & 309.000000 & 128.000000 & 11.000000 & 3.000000 \\
        Max. Nodes & 516.000000 & 451.000000 & 489.000000 & 488.000000 & 520.000000 & 328.000000 & 163.000000 & 35.000000 \\
        Min. Density & 0.009727 & 0.010006 & 0.006463 & 0.007692 & 0.008856 & 0.011315 & 0.040532 & 0.095385 \\
        Max. Density & 0.029232 & 0.051309 & 0.013846 & 0.014834 & 0.013633 & 0.023745 & 0.894727 & 1.000000 \\
        Min. Coms & 11.000000 & 8.000000 & 13.000000 & 12.000000 & 13.000000 & 10.000000 & 1.000000 & 1.000000 \\
        Max. Coms & 20.000000 & 18.000000 & 20.000000 & 17.000000 & 18.000000 & 17.000000 & 10.000000 & 6.000000 \\
        \bottomrule
    \end{tabular}
\end{table*}

\sisetup{round-mode=figures,round-precision=3,scientific-notation = false , table-format=1.5}
\begin{table*}[!htb]
    \centering
    \caption{MMD scores for R-MAT and GRAN on each dataset summarised in Table~\ref{tab:datasets}. Lower is better;  bold text indicates a superior result.}
    \label{tab:results_mmd}    
    \begin{tabular}{llSSSSSSSS}
    \toprule
    Dataset    &    Model            &  \rx{MMD\fsub{Nodes}}           &  \rx{MMD\fsub{Degree}}      &  \rx{MMD\fsub{Clustering}}   &  \rx{MMD\fsub{Spectral}}      &  \rx{MMD\fsub{Steps}}   &  \rx{MMD\fsub{Saturation}}    &  \rx{MMD\fsub{Paths}} &  \rx{MMD\fsub{Louvain}}    \\\midrule
    \multirow{2}{*}{Facebook} & GRAN                  &  \fbf 0.0280         & \fbf 0.00689     &  \fbf 0.0949      &  \fbf 0.00479      &  \fbf 0.00163    &  \fbf 0.00252          &  \fbf 0.0200        &  \fbf 0.0305    \\
             & RMAT                  &  0.110                &  0.0265           &  0.604             &  0.0411             &  0.00432          &  0.00435                &  0.103               &  0.0622          \\\midrule
    
    \multirow{2}{*}{GitHub} & GRAN                    &  0.0928               & \fbf 0.0164      &  \fbf 0.216       &  \fbf 0.0133       & \fbf 0.00318     &  \fbf 0.00375          &  \fbf 0.0367        &  \fbf 0.157     \\
           & RMAT                    &  \fbf 0.0352         &  0.0634           &  1.170              &  0.0613             & 0.0152            &  0.0131                 &  0.0735              &  0.210           \\\midrule
    
    \multirow{2}{*}{Twitch} & GRAN                    &  0.0591               & \fbf 0.00543     &  \fbf 0.136       &  \fbf 0.00854      & \fbf 0.00644     &  \fbf 0.00859          &  \fbf 0.0239        &  \fbf 0.0832    \\
           & RMAT                    &  \fbf 0.0213         &  0.0185           &  1.140              &  0.0246             & 0.00925           &  0.00900                &  0.0833              &  0.181           \\\midrule
    
    Deezer & GRAN                & 0.133                 &  0.0360     &  \fbf 0.0557      &  \fbf 0.0197       & \fbf 0.0481      &  \fbf 0.0499           &  \fbf 0.00768       &  0.182           \\
    Ego    & RMAT                & \fbf 0.0145          &  0.0359     &  0.0875            &  0.0756             & 0.0787            &  0.0854                 &  0.248               &  \fbf 0.158     \\
    \bottomrule
    \end{tabular}

\end{table*}


The aim of this section, and of Section~\ref{sec:vis-inspec}, is to answer \ref{rq1} and \ref{rq2}. To this end we empirically evaluate the quality of generated topologies using the testing set of each dataset. We generate 40 networks using GRAN for each sampled dataset (Facebook, Twitch, Github) to match the size of each test set. We limit the number of generated networks for the Deezer dataset to 800 to allow computation of metrics in reasonable time. We compute metrics using a random 800 network sample of the Deezer test-set.

GRAN demonstrated a tendency to produce separate connected components, so for measurement, we take the largest of these. This means that the number of nodes in the samples actually used for measurement is often less than the number found in the training or test networks. The same measurements in Table~\ref{tab:datasets} are computed for the test and generated sets, and shown in Table~\ref{tab:gen_stats}.

GRAN models fit on networks with areas of higher sparsity (Facebook, Twitch) display a higher tendency to produce separate connected components, hence the significantly lower maximum and minimum node numbers compared to their test sets. This reduction in number of nodes likely also explains their higher maximum densities compared to their test sets, as larger networks tend to have more sparse regions, leading to lower densities. In contrast, for the more consistent GitHub networks, GRAN is much more closely able to match numbers of nodes, and has a far lower tendency to produce separate components. We discuss this further in Section~\ref{sec:con_comp}.

Fit on our largest dataset, the Deezer networks, GRAN produces far smaller networks. The two distributions of network sizes peak at the same point, but the GRAN distribution is far shorter-tailed than that for real networks. GRAN produces some especially small connected components, which are actually smaller than any present in its training data.


Attempts to fit R-MAT networks on each network in our testing set resulted in some failures. This is due to the nature of how we fit and generate R-MAT networks using Networkit. We fit parameters~$a,b,c,d$ on the adjacency matrix of a given test set network, with each representing the proportion of edges in their respective quadrant of the matrix. In some cases or node orderings one quadrant may be ``empty'' - i.e. containing no edges. In this case fitting and generation will fail. This would in practise be avoided by researchers working extensively with R-MAT, and in our situation, is un-common enough not to alter results significantly. Roughly $4\%$ of attempted R-MAT fitting and generation failed in a given dataset.

Having generated R-MAT networks, we compute the previously detailed MMD scores, for both GRAN and R-MAT generated networks, using the test set for each dataset. These results are detailed in Table~\ref{tab:results_mmd}.

With the exception of the Facebook dataset, as expected, MMD\fsub{Nodes} is lower for R-MAT. This is unsurprising, as each R-MAT network is specifically fits the number of nodes in each test set network. Without R-MAT failures, which are more common for the Facebook data, we'd expect MMD\fsub{Nodes}$=0$ for any R-MAT generated set of networks.

For sampled datasets (Facebook, GitHub, Twitch) GRAN outperforms R-MAT. This is particularly true of structural metrics, such as MMD\fsub{Clustering}, MMD\fsub{Spectral} or MMD\fsub{Saturation}, where MMD scores for GRAN are generally much lower than that for R-MAT networks. This might indicate that GRAN more accurately reproduces the more complex aspects of social network topologies.

For GRAN and R-MAT networks generated based on Deezer networks results are less decisive. As with GitHub and Twitch, node scores are superior for R-MAT, but to a far greater degree, with almost a 10 fold difference in MMD\fsub{Nodes}. In fact, MMD\fsub{Nodes} for R-MAT is the lowest for any dataset, possibly due to the lower failure rate on smaller networks and larger dataset sizes. 

Structural scores are still superior for GRAN, but to a smaller degree than on the other datasets. The exception here is shortest paths, where the discrepancy in MMD\fsub{Paths} between GRAN and R-MAT is highest across any dataset we use. This may be due to R-MAT's inability to represent the ``central node'' nature of an ego network. The longest possible path across \ftx{any} ego network is two edges, but if R-MAT is unable to capture that ego networks are built around a central node, paths are longer. 


\subsubsection{Quality of Topologies - Visual}
\label{sec:vis-inspec}

\begin{figure}[!hbt]
    \centering
    \includegraphics[height = \linewidth, angle = -90]{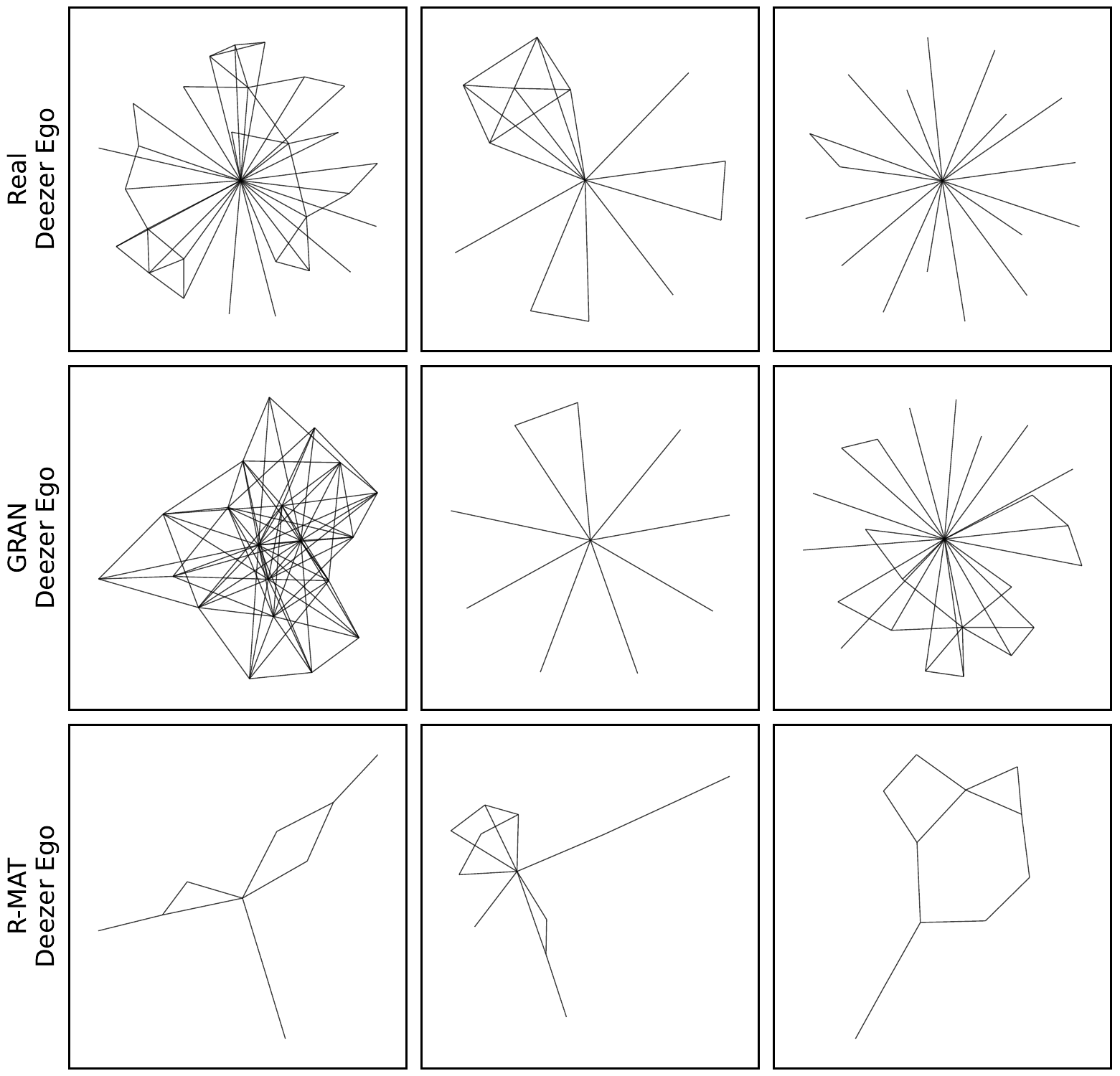}
    \caption{Comparison of real networks and synthetic networks produced by GRAN and R-MAT trained on the Deezer dataset detailed in Table~\ref{tab:datasets}.}
    \label{fig:vis_deezer}
\end{figure}

\begin{figure}[!hbt]
    \centering
    \includegraphics[height = \linewidth, angle = -90]{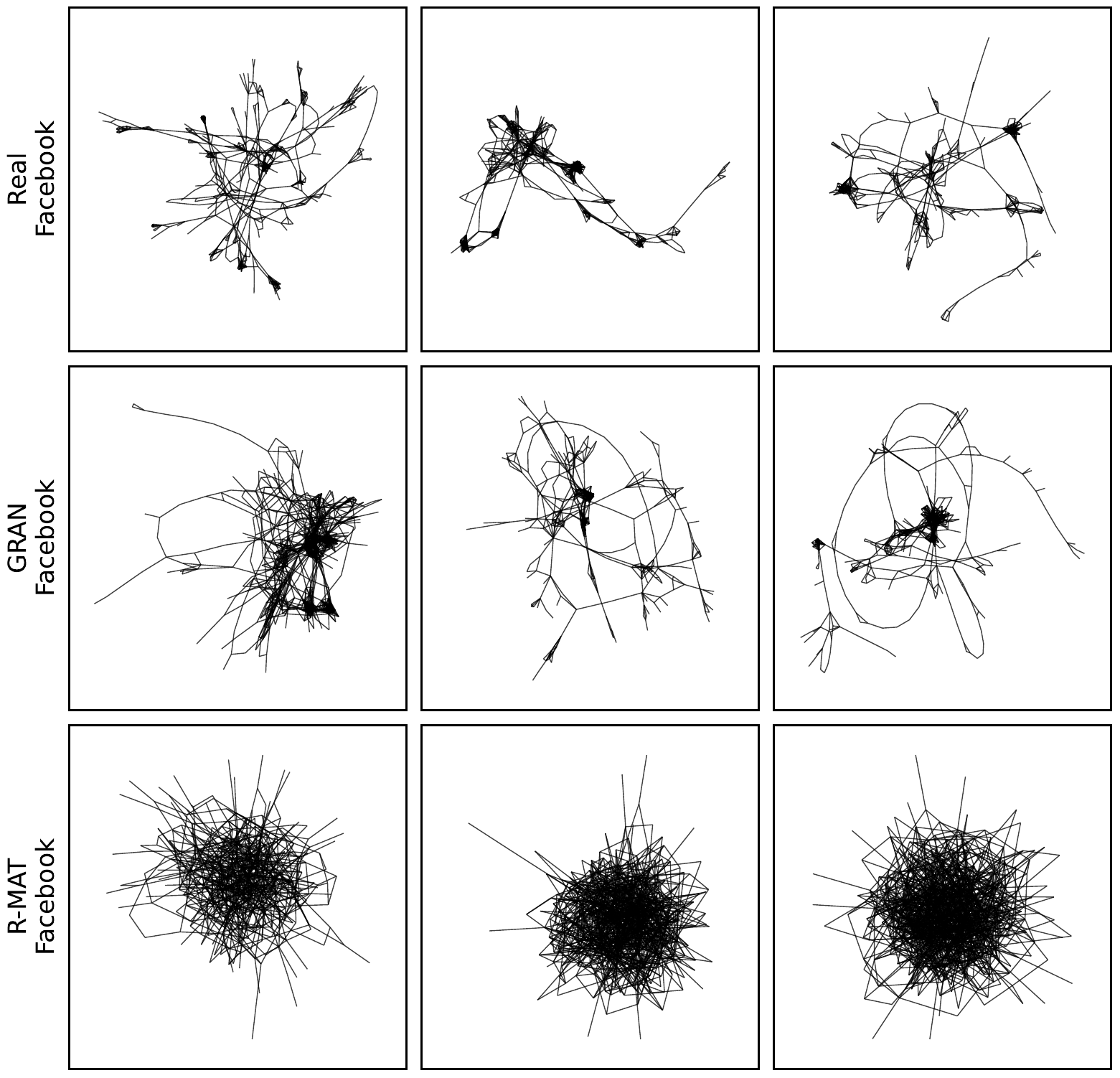}
    \caption{Comparison of real networks and synthetic networks produced by GRAN and R-MAT trained on the Facebook dataset detailed in Table~\ref{tab:datasets}.}
    \label{fig:vis_fb}
\end{figure}

\begin{figure}[!htb]
    \centering
    \includegraphics[height = \linewidth, angle = -90]{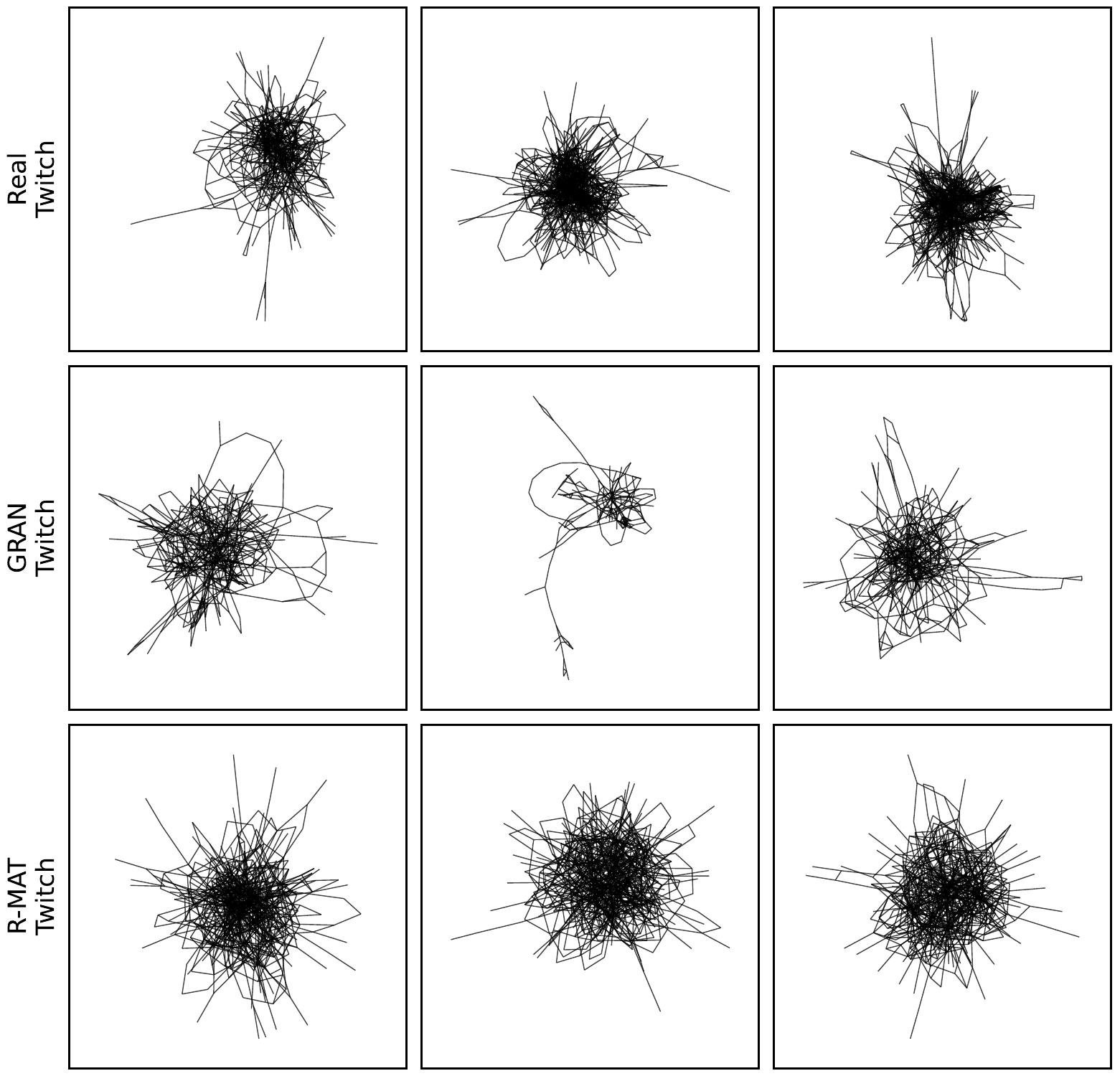}
    \caption{Comparison of real networks and synthetic networks produced by GRAN and R-MAT trained on the Twitch dataset detailed in Table~\ref{tab:datasets}.}
    \label{fig:vis_twitch}
\end{figure}

\begin{figure}[!ht]
    \centering
    \includegraphics[height = \linewidth, angle = -90]{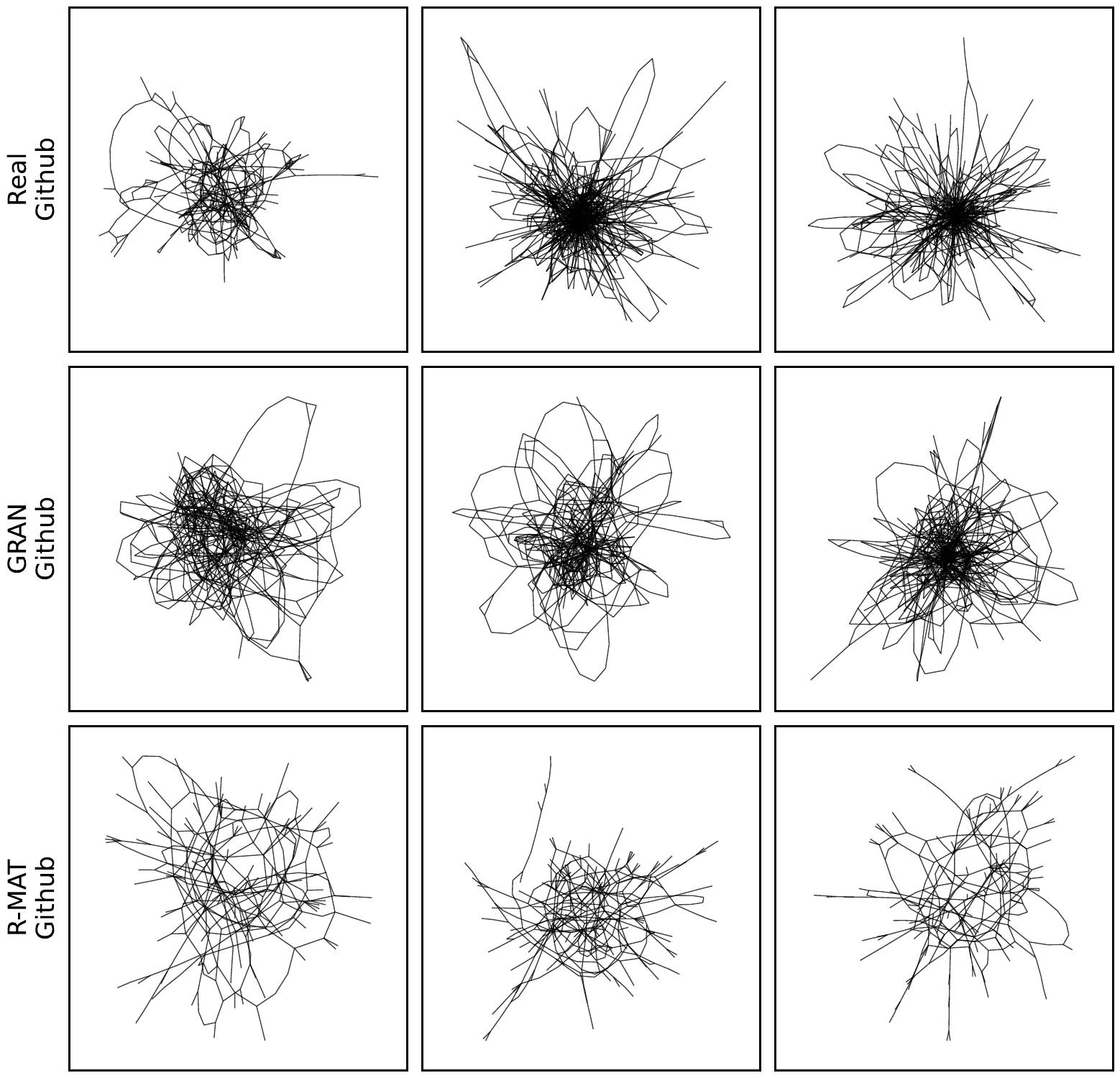}
    \caption{Comparison of real networks and synthetic networks produced by GRAN and R-MAT trained on the Github dataset detailed in Table~\ref{tab:datasets}.}
    \label{fig:vis_git}
\end{figure}

In this section we show generated networks from R-MAT and GRAN alongside their real counterparts. We show only the largest connected components resulting from each generative method, and the reader should note that the R-MAT networks are not fit on the test-set networks displayed alongside them. In Fig.s~\ref{fig:vis_deezer}~and~\ref{fig:vis_fb} we show networks related to the Deezer and Facebook datasets, and in Fig.s~\ref{fig:vis_twitch}~and~\ref{fig:vis_git} we show networks related to the Twitch and GitHub datasets. Layouts are via networkx.spring\_layout, a force-based algorithm, with 100 iterations of simulation, and for GRAN-generated networks, only the largest connected component is displayed.

Visualisation of the Deezer networks in Fig.~\ref{fig:vis_deezer} corroborates the findings of Section~\ref{sec:empirical}. The real ego networks have, crucially, one central node with which all other nodes share an edge. The real networks also have connections between non-central nodes, which in some cases are actually a complete network including the central node. GRAN has properly re-created these features, albeit with fewer nodes than in the displayed test set examples.

R-MAT generated networks for the Deezer dataset clearly do not have a central node. The degree distribution is close to correct, with one or two high-degree nodes and many low-degree nodes, but networks have a molecule-like structure, compared to the hub-and-spoke like topologies in the test set or generated by GRAN.

The real Facebook networks in Fig.~\ref{fig:vis_fb} show clusters and sparsity in-between. This is characterised visually by what we term ``arcs'', where nodes have two neighbours, leading to strings of nodes with no further structure. This is shown in the GRAN generated networks and to a lesser extent in the R-MAT networks. However, the R-MAT networks seem essentially to be one large cluster of nodes, without any complex structure.
The same is true to some extent of the Twitch networks in Fig.~\ref{fig:vis_twitch}. The real Twitch networks do not show the same level of structure as with the Facebook networks, but do still exhibit arcs emerging from the principle mass of nodes. Again, R-MAT generated networks do not show these same structural dynamics.

The trend continues with GitHub networks in Fig.~\ref{fig:vis_git}, where arcs are generated by GRAN, but not by R-MAT. Again GitHub networks don't have the same structural dynamics as Facebook networks, with only one central grouping, more closely resembling the Twitch networks.



\subsubsection{Run Time \& Computational Cost}


This section addresses \ref{rq3}. Firstly, as R-MAT does not as-such have a ``training time'', we list the times in training for GRAN on each of our datasets in Table~\ref{tab:gentime}. Variation between the Facebook, Twitch and GitHub datasets is likely due to the varying densities of each set of networks. This varying density changes the in-memory size of each batch, and as GRAN employs an MPNN, will also change the cost of the message-passing stage. We also detail in Table~\ref{tab:gentime} the mean time taken by R-MAT and GRAN to generate a single network. For R-MAT this also involves fitting to a random network from the test set for each dataset.



\begin{table}[hbt]
    \caption{Training times for GRAN on each dataset, along with the mean time taken for GRAN and R-MAT to generate a network.}
    \label{tab:gentime}
    \centering
    \begin{tabular}{lrrr}\toprule
                   &    {Training}      & \multicolumn{2}{r}{Generation (seconds)}  \\
                   &    {(hr:min:sec)}  &  {GRAN}        &  {R-MAT} \\\midrule
    {Facebook}     &    12:01:07        &  1.05          &  0.00501\\
    {Twitch}       &    16:36:27        &  0.45          &  0.00392\\
    {GitHub}       &    16:09:45        &  0.525         &  0.00338\\
    {Deezer}       &    41:51:06        &  0.775         &  0.000375\\\bottomrule
    \end{tabular}

\end{table}


From Table~\ref{tab:gentime} we can say that GRAN is, to a large extent, more of a time and computational investment than R-MAT. The time investment is not, however, massive. At a maximum generation time of one second, a dataset of nearly a hundred thousand networks can be created in as little as 24 hours. GRAN also has the advantage of being able to generate more networks than were in the original dataset with realistic but, due to its stochastic procedure, potentially unique structure in the topology of the resulting networks. This represents a common compromise for those selecting a model: expressivity vs computational cost.

\section{Discussion}
\label{sec:discussion}

This section details first adresses the results and findings of our experiments. We then detail the assumptions and risks to validity that should be taken into account when considering these findings.

\subsection{Findings}

We structure our findings according to the takeaways from our research questions.
\subsubsection{\ref{rq1}: How could we generate synthetic social network topologies comparable to real networks?} 

We find that realistic synthetic social networks can be produced by GRAN. The networks produced correspond closely to their real counterparts, including when measured using SIR simulations and Louvain community detection. However GRAN shows a tendency to produce multiple connected components. The largest of these can be significantly smaller than those in the training data. 

We hypothesise that the size of network samples should be topologically significant, instead of simply sampling a given number of nodes.
This may allow GRAN to more properly encode network size, and might alleviate the issues we find with multiple connected components.

\subsubsection{\ref{rq2}: When is a GNN model more useful for generating synthetic datasets than the current standard?}

We are able to compare the networks generated by R-MAT, a rule-based method, and by GRAN, a GNN model. We use this comparison to evaluate the compromise researchers might face when choosing between types of model. 

R-MAT generated networks avoid the issues with multiple connected components that we find with GRAN, but on all other metrics, GRAN performs better than R-MAT. This is especially true for structural and social network specific metrics, where GRAN strongly out-performs R-MAT. In particular the existing issues with rule-based models are clear in networks generated based on the Deezer networks. Here R-MAT is unable to reproduce the essential nature of an ego network, whereas the more expressive GRAN produces realistic ego networks, as demonstrated by both visualisations and the measurement of shortest paths.

We conclude that R-MAT can be used where reasonable but not realistic approximations of social networks are needed, for example in applications where power-law degree distributions are paramount, or in rapid early testing of models. In situations where realism is essential GRAN should be employed. This is especially true for those wishing to create synthetic datasets for distribution.

\subsubsection{\ref{rq3}: What is the trade-off between computational cost and the ability to create realistic social networks?}

Our experiments show that R-MAT is significantly less costly to use than GRAN, especially when including GRAN's training time, which would likely hold true for other rule-based methods. The greater expressivity and complexity of GRAN means that it is more able to represent complex dynamics in networks, but requires much more computational and time investment, as well as a substantial training dataset. Again this would likely hold true of other deep-learning/rule-based method comparisons, which we leave as an area for future research.


We do find that GRAN is not hugely costly within the context of deep-learning models. With good results over a training time of less than a day, and generation time-per-network of at most around one second, constructing datasets of synthetic networks is feasible. We did not make use of HPC facilities and we envisage superior performance if GRAN is trained more extensively.


\subsection{Assumptions and Threats to Validity}

Here we detail assumptions we've made, potential limitations that might represent threats to the validity of our findings, and areas on which future research might focus.
\subsubsection{Sampling Procedure}

In using ESWR we assume that an exploration sampler is able to take representative samples. \citet{Leskovec2006SamplingGraphs} find that exploration samplers perform well in this task, but also note that there is a lower limit to the size of a sample, which they set at around 15\%. Our sub-network samples are (on average) below this size, which was in order to remain tractable as training samples for GRAN. This means that, while possibly remaining representative samples of the larger networks, our sampled sub-networks may not be adequately representative.

\subsubsection{GRAN Generated Graph Size}
\label{sec:con_comp}

A generative deep-learning model like GRAN aims to encode and then reproduce a high-dimensional representation of its training data.
GRAN adds nodes in blocks, then uses an MPNN to determine which nodes already in the network should share edges with the newly added nodes.
GRAN can determine that a new node \ftx{shouldn't share any edges} with the existing network, in which case it is a new component of size one. GRAN runs through to the maximum number of nodes in a network in this manner, with smaller networks created by leaving the extra nodes un-connected.
The potential issue with our datasets is that the size of a network is determined not by network structure, but as an input parameter, which which may not be encoded through topology.

We argue this may be why GRAN was found to generate multiple un-connected components.
If a single node mid-way through the network building process is un-connected, new nodes can be connected either to the existing network or to this new separate component, and successive nodes could be more likely to be attached to the new component as it grows.
A potential remedy might be to sample sub-networks not according to number of nodes but instead according to some other network measure, for example communities or number of edges.
These might be better encoded by GRAN or other deep-learning models.


\subsubsection{Use of R-MAT}

We have assumed that R-MAT, due to its popularity, is a fair model against which to compare a deep-learning method like GRAN. However, R-MAT is not an especially recent model, and likely remains popular due to its simplicity and ability to generate realistic power-law degree distributions  \cite{Nettleton2021MEDICI:Generator, Nettleton2016AGraphs}. A more recent model which itself has been used as a benchmark is \citet{Wang2011LeveragingSystems}'s generator, which given its stated ability to generate community structures, may for future research may be a more fair baseline.

\subsubsection{GRAN training}

A significant factor to consider is that we do not aim to produce the best possible performance with GRAN, and so we have not performed any hyper-parameter optimisation process, and trained only for a comparatively short time. \citet{Liao2019EfficientNetworks} do not give details of training duration, number of epochs, batch sizes and so on. Based on configuration files packaged with GRAN, specifically for the DB point cloud dataset from \citet{Neumann2013GraphGrasping}, they train GRAN for 20k epochs on a dataset of 41 much larger networks than we use. These networks, while larger than ours, are $k$ nearest neighbour networks. This means that comparatively they are arguably less topologically complex. We suspect that training GRAN for a longer duration---as appears to have been the case with the original \citet{Liao2019EfficientNetworks} work---would bring greater performance.







\section{Conclusion}
\label{sec:conclusion}

We evaluate the ability of recent deep Graph Neural Network (GNN) models in generating synthetic social networks. Part of this method allowed comparison with a rule-based model, which are currently more widely used for this task. As a GNN model we use the Graph Recurrent Attention Network  (GRAN) \cite{Liao2019EfficientNetworks}, an auto-regressive deep-learning model, which has in other fields demonstrated its ability to produce semantically meaningful structure in generated networks. For a benchmark comparison, as a rule-based method, we use the Recursive-MATrix (R-MAT) model from \citet{Chakrabarti2004R-MAT:Mining}.

We first sample sub-networks from three larger datasets sourced from \citet{Rozemberczki2019GemSec:Clustering} via Exploration Sampling With Replacement (ESWR). We also include as a fourth dataset the Deezer networks from \citet{Rozemberczki2020KarateGraphs}, as given ego networks are typically small and possess the defining feature of a central node, ego networks represent a useful toy dataset.

We employ GRAN on 80:20 train:test splits of each dataset, use the trained models to construct datasets of synthetic networks, then compute MMD scores against the test set for R-MAT and GRAN. MMD scores are calculated with network metrics as well as through application of social network specific measurements.

GRAN demonstrated a propensity to generate multiple connected components in networks with regions of sparsity. This leads to MMD\fsub{Nodes} being superior for R-MAT, as each R-MAT network is fit specifically to a network in the test set, whereas GRAN ``loses'' nodes to these smaller connected components. A remedy to this issue with GRAN might be to use a more topologically consistent sampling method. As we simply employ an exploration sampler up until a given number of nodes are sampled, the size of the resulting network is not necessarily encode-able by GNN models. A sampling method with termination based a topologically meaningful metric, employed in the same manner as we use in this work, is left as an area for future research.


GRAN, on all other MMD scores, significantly out-performs R-MAT. This is particularly true of structural or social-network specific measurements such as shortest path lengths, where complex topological structure plays a larger part. In the Deezer networks, GRAN demonstrates an ability to encode the essential ``one central node'' nature of an ego network, whereas R-MAT is unable to. Characteristics of each network dataset we've constructed are reproduced better by GRAN than by R-MAT.

Computational cost is significantly higher for GRAN than R-MAT, but not so high as to be un-feasible for researchers, or those with access to moderate computing power. We conclude that R-MAT could be used for rapid model testing, or in applications where a power-law degree distribution is paramount, and that GRAN should be employed in applications where network realism is worth the moderate computational cost. In particular this would be true for those constructing synthetic datasets for tasks that could include topology as a feature.



\aod{Cut "online" parts from bib - don't need country - remove links from refs - Check everything is "proceedings of" - }

\end{document}